# A Tale of Two Desolvation Potentials: An Investigation of Protein Behavior Under High Hydrostatic Pressure

*Andrei G. Gasic*[†,‡] *& Margaret S. Cheung*[†,‡,] *

[†]Department of Physics, University of Houston, Houston, Texas, 77204, USA

[‡]Center for Theoretical Biological Physics, Rice University, Houston, Texas, 77005, USA

*Corresponding author email: mscheung@uh.edu

ABSTRACT

Hydrostatic pressure is a common perturbation to probe the conformations of proteins. There are two common forms of pressure-dependent potentials of mean force (PMFs) derived from hydrophobic molecules available for the coarse-grained molecular simulations of protein folding and unfolding under hydrostatic pressure. Although both PMF includes a desolvation barrier separating the well of a direct contact and the well of a solvent-mediated contact, how these features vary with hydrostatic pressure is still debated. There is a need of a systematic comparison of these two PMFs on a protein. We investigated the two different pressure-dependencies on the desolvation potential in a structure-based protein model using coarse-grained molecular simulations. We compared them to the known behavior a real protein based on experimental evidence. We showed that the protein's folding transition curve on the pressure-temperature phase diagram depends on the relationship between the potential well minima and pressure. For protein that reduces the total volume under pressure, it is essential for the PMF to carry the feature that the direct contact well is essential less stable than the water-mediated contact well at high pressure. We also comment on the practicality and importance of structure-based minimalist models for understanding the phenomenological behavior of a protein under a wide range of phase space.

**Introduction**

Not only do proteins unfold under high heat, but also in the presence of high hydrostatic pressure. This effect has been known since the early 1900's,[1,2] and the equation of state on the pressure-temperature plane was formalized by Hawley in 1971.[3] While protein unfolding by heat is more intuitive, pressure denaturation can by explained from Le Chatelier's principle, in which pressure unfolds proteins due to a negative volume change. The molecular origin of this negative volume change was recently discovered to be the penetration of water into the hydrophobic core, causing loss of the protein's cavities.[4,5]

Computational simulations are essentially to gain further insight into specific pressure perturbed folding mechanisms. Pressure denaturation of proteins has been studied by all-atom molecular dynamics simulation,[6-9] but it is computationally costly since pressure unfolds proteins on a longer time scale than that of heat denaturation,[10] often requiring sampling tricks.[11] An alternative to all-atom models are structure-based coarse-grained models. Structure-based models render an energy landscape with minimal frustration and contains a funneled landscape with a dominant basin of attraction, corresponding to an experimentally determined configuration.[12,13] The mechanism that drives protein folding from unfolded conformations to few unique conformations where the hydrophobic residues coalescences to a "hydrophobic core" is similar to what drives oil separating from water. As such, these models are computationally inexpensive, allowing long-timescale simulations even for large proteins and complex systems. Similar to how the "Ising model" is used to develop the general theories of phase transitions, or how the "ideal gas" illustrates the basic notions of fluid behavior, structure-based models of proteins are used to understand fundamental aspects of protein folding and dynamics.[13,14] Therefore, to understand large systems, such as protein folding *in vivo*,[15,16] structure-based minimalist models are essential for developing new theories.

Pressure-dependent hydrophobic interactions in coarse grained protein models are approximated by using the potential of mean force (PMF) between two methane molecules at various hydrostatic pressures,[17-20] since methane molecules are a simple model for the interaction of hydrophobic residues in a protein. Understanding the behavior of two contacting methane molecules in aqueous solution provides insight into the mechanism of hydrophobic collapse during protein folding. In general, this PMF contains a contact well, a solvent-mediated well, and a desolvation barrier between the two wells, as shown in Figure 2. The presence of the desolvation barrier is due to the free energy cost for two methane molecules to penetrate into the first hydration shell from the well of a water-mediated contact before they reach another well of a direct contact. Pressure ($P$) and temperature ($T$) can cause changes in the depths of wells and height of the barrier.

However, the exact description of the PMF are still debated between two methane molecules[18,21-24] and in proteins.[25-29] Hummer *et al.*[18] find a pressure-dependent PMF with a contact well energy that weakens favoring the water-mediated well as pressure increases. Opposing Hummer *et al.*,[18] Dias and Chan show that the contact well deepens as pressure increases.[19] Because of the different pressure-dependencies of these two potentials, they produce different folding phase behavior of proteins on the *P-T* plane. Here, we investigate these two different pressure-dependent desolvation potentials and compare it to the known behavior of a well-studied protein under high hydrostatic pressure, phosphoglycerate kinase (PGK), based on experimental evidence.[30–32] Before presenting the results, we give an overview of pressure-denaturation relevant to the current investigation.

## Overview of Pressure-Denaturation

**Folding thermodynamics in the *P-T* phase diagram.** Hawley's theory dictates that the contributions of the changes in entropy ($\Delta S \equiv S_U - S_F$) and volume ($\Delta V \equiv V_U - V_F$) between unfold (*U*) and folded (*F*) protein phases result in an elliptical-shaped coexistence ($\Delta G = 0$) curve in *P-T* space (Figure 1).[3] From the Clausius-Clapeyron relation, the slope of the coexistence curve between the two phases is,

$$\left.\frac{dP}{dT}\right|_{\Delta G=0} = \frac{S_U - S_F}{V_U - V_F} = \frac{\Delta S}{\Delta V}. \qquad (1)$$

Since the coexistence curve is elliptical, both $\Delta V$ and $\Delta S$ can be positive, negative, or zero depending on the *P* and *T*. Figure 1 contains two different examples of elliptical curves to help breakdown the contributions of entropy and volume in the discussions to follow in this section and in the Discussion section.

From Le Chatelier's principle, pressure shifts the equilibrium toward the phase with the smallest volume to minimize the free energy. Therefore, for hydrostatic pressure to unfold a protein, the partial molar volume of the unfolded phase ($V_U$) must be smaller than that of the folded phase ($V_F$); i.e., the change in volume must be negative ($\Delta V < 0$).[33,34]

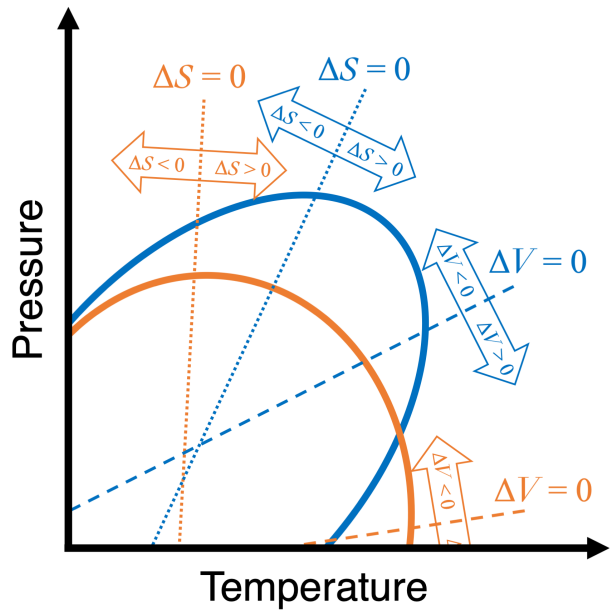

**Figure 1.** Schematic *P-T* phase diagrams for two proteins with different elliptical coexistence ($\Delta G = 0$) curves. The dotted lines represent $\Delta S = 0$ (left of the line, $\Delta S$ is negative; right of the line, $\Delta S$ is positive), and the dashed lines represent $\Delta V = 0$ (above of the line, $\Delta V$ is negative; below of the line, $\Delta V$ is positive). The two ellipses differ by a shift in the pressure direction and a shear of the $\Delta V$ and $\Delta S$ axes. The origin of the plot is at 0.1 MPa (1 atm) and 273 K. For the orange curve, $\Delta V$ is mainly negative for experimentally relevant *P* and *T*. Whereas, the blue curve has a large region where $\Delta V$ is positive. Examples of proteins with a similar elliptical coexistence curves is ribonuclease (orange) and chymotrypsinogen (blue).

In general, $\Delta V$ is negative at room temperature;[35] however, positive $\Delta V$ can be observed, specifically at high temperature[3] or for alpha-helix peptides.[36,37] One of the key factors in producing the elliptical phase diagram for protein folding stability in the *P-T* plane is the temperature dependence of $\Delta V$. Formally, $\Delta V$ is given as

$$\Delta V = \int_{T_0}^{T} \Delta\alpha \, dT' + \int_{P_0}^{P} \Delta\kappa \, dP' = \Delta V_0 + \Delta\alpha(T - T_0) + \Delta\kappa(P - P_0), \qquad (2)$$

where the change in thermal expansivity is $\Delta\alpha$, the change in compressibility is $\Delta\kappa$, and $\Delta V_0 \equiv \Delta V(T_0, P_0)$. Usually, the magnitude of $\Delta\kappa$ is small, giving an almost negligible pressure-dependence to $\Delta V$. At room temperature, generally $\Delta\kappa < 0$. Additionally, $\Delta\alpha$ tends to be greater than zero, which may be due to the greater degree of hydration of unfolded proteins.[38]

The $P$ and $T$ dependencies of $\Delta S$ is the other factor that contributes to the elliptical phase diagram. Since a folded protein has a lower entropy than that of an unfolded protein, heat shifts the population to the phase with highest entropy. At low temperature, the low entropy state of the solvent is favored, which is reconciled by unfolding the protein.[39,40] Note that in this investigation, we will not consider cold-denaturation.

**Molecular mechanism of pressure unfolding.** The change in partial molar volume stems from changes in void (or solvent inaccessible cavity), van der Waals, and hydration volumes:[41]

$$\Delta V = \Delta V_{\text{vdW}} + \Delta V_{\text{void}} + \Delta V_{\text{hyd}} \qquad (3)$$

However, $\Delta V_{\text{vdW}} \approx 0$ for most cases, which means $\Delta V_{\text{vdW}}$ can be ignored.

For partial molar volume changes in void, $\Delta V_{\text{void}}$, since folded proteins are not perfectly packed, dry voids (or cavities) of varying sizes and shapes are distributed heterogeneously throughout a protein.[42,43] High pressure induces unfolding by introducing solvent into the protein structures, eliminating the cavities and decreasing the overall solvent accessible volume.[45] Water penetrates into the hydrophobic core of proteins because of the reduced solvent-solute interfacial free energy when pressure increases.[44]

For partial molecular volume changes in hydration, $\Delta V_{\text{hyd}}$, hydration of newly solvent-exposed side chains upon unfolding also contributes to change in partial molar volume by changing the density of the hydration layer. This effect depends on the chemical properties of the side chains and the topography the surface area. Usually, $\Delta V_{\text{hyd}} > 0$ and is small in magnitude, which may be due to polar and apolar hydration influences cancelling each other. Thus, the decrease in void volume, $\Delta V_{\text{void}}$, overcomes the increased hydration volume, $\Delta V_{\text{hyd}}$, in order for pressure-denaturation to occur.[45] Regardless of the hydration effect, the key mechanism that unfolds a protein under high pressure is the elimination of solvent excluded cavities because of water penetrating the hydrophobic core; therefore, $\Delta V < 0$ to unfold a protein under high pressure.

## Theoretical Model and Simulation Methods

**Structure-based coarse-grained model.** Our simulations use a structure-based coarse-grained model, which is minimalist protein model ("beads on a chain") that incorporates experimentally-derived structural information.[46] Structure-based models are the "ideal gas" model of protein folding, and are used for investigating of a wide range of folding mechanisms.[14,47] Since pressure unfolds proteins at an order of magnitude (or more) slower than heat unfolding, long-timescale simulations are also crucial for high-pressure unfolding. Therefore, structure-based, minimalist-model simulations provide statistically significant results. Analogous to adding specific complexity to the ideal gas model to study specific interactions, we add the desolvation potential[48] to the native interactions of a protein to account for the free energy cost of expelling a water molecule in the first hydration shell between two hydrophobic residues.[49] This desolvation model predicts a folding mechanism based on water expulsion from the hydrophobic core, which has been observed by all-atomistic molecular dynamics[50] and validated by experiments in which the volume or polarity of amino acids is changed by mutation.[51]

The Hamiltonian of this structure-based model is as follows:

$$\mathcal{H}(\Gamma, \Gamma^0) = \sum_{i<j} K_r (r_{ij} - r_{ij}^0)^2 \delta_{j,i+1} + \sum_{i \in \text{angles}} K_\theta (\theta_i - \theta_i^0)^2$$

$$+ \sum_{i \in \text{dihedrals}} K_\phi \left( \{1 - \cos[\phi_i - \phi_i^0]\} + \frac{1}{2} \{1 - \cos[3(\phi_i - \phi_i^0)]\} \right)$$

$$+ \sum_{\substack{i,j \in \text{native} \\ |i-j|>4}} U(r_{ij}, \epsilon, \epsilon', \epsilon'') + \sum_{i,j \notin \text{native}} \epsilon_0 \left( \frac{\sigma}{r_{ij}} \right)^{12}, \quad (4)$$

where $\Gamma$ is a configuration of the set $r, \theta, \phi$. The $r_{ij}$ term is the distance between $i^{th}$ and $j^{th}$ residues, $\theta$ is the angle between three consecutive beads, and $\phi$ is the dihedral angle defined over four sequential residues. $\delta$ is the Kronecker delta function. The native state values of $r^0, \theta^0, \phi^0$ were obtained from their crystal structure configuration, $\Gamma^0$, which $= \{\{r^0\}, \{\theta^0\}, \{\phi^0\}\}$. In the backbone terms, $K_r$, $K_\theta$, and $K_\phi$ are force constants of the bond, bond-angle, and dihedral potentials, respectively. We used $K_r = 100\epsilon_0$, $K_\theta = 20\epsilon_0$, and $K_\phi = \epsilon_0$, where $\epsilon_0$ ($\equiv 1$) is the solvent averaged energy at ambient pressure.

The native contact interactions are governed by the desolvation potential, $U(r)$ (shown in Figure 2(b)), which depends on the distances between two contacts, $r$, the depth of the first well, $\epsilon$, the depth of the water-mediated well, $\epsilon'$, and the height of the desolvation barrier, $\epsilon''$. Note that $U(r)$ is only between residues with an index separation greater than four. The potential is modified to incorporate the effects of high-pressure by giving pressure-dependencies to $\epsilon$, $\epsilon'$, and $\epsilon''$.

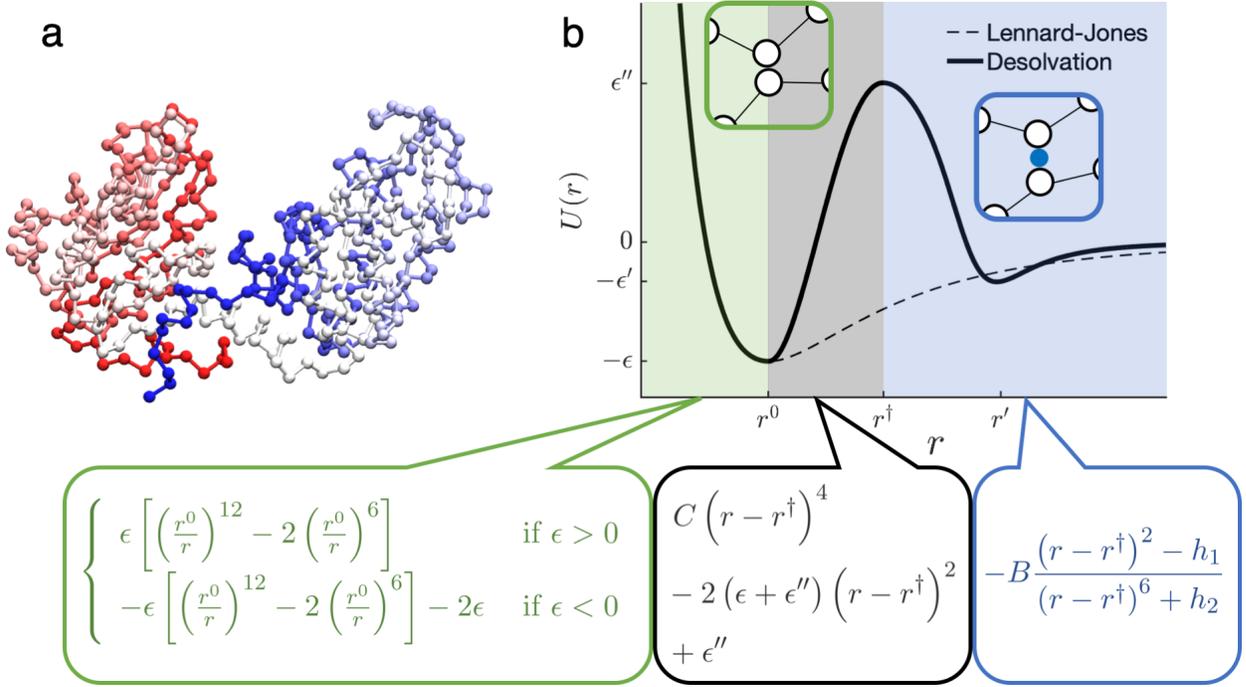

**Figure 2.** (a) A coarse-grained representation of PGK (PDB ID: 1QPG). Each amino acid is coarse-grained to a single bead. The N- and C-terminus domains are in red and blue, respectively. (b) The desolvation potential (solid line) compared to the Lennard-Jones potential (dashed line). Here, $r^0, r^\dagger$, and $r'$ are the positions of the minimum of the first well, maximum of the desolvation barrier, and the minimum of the second well, respectively. The separation between $r^0$ and $r'$ is the size of a single water molecule, 0.8 σ, and $r^\dagger = (r^0 + r')/2$. The piece-wise terms of the desolvation potential are shown below pointing to their corresponding colored sections, where constants $C = \frac{(\epsilon+\epsilon'')}{(r^\dagger-r^0)^2}, B = 3\epsilon'(r'-r^\dagger)^4, h_1 = \frac{2}{3}\frac{(r'-r^\dagger)^2}{\epsilon'/\epsilon''+1}$, and $h_2 = 2\frac{(r'-r^\dagger)^6}{\epsilon''/\epsilon'+1}$. The green section $(r < r^0)$ is the Lennard-Jones potential for positive or negative $\epsilon$. The attractive part of the contact well is highlighted in gray $(r^0 \leq r < r^\dagger)$, and the water-mediated minimum is highlighted in blue $(r^\dagger \leq r)$.

**Pressure-dependent desolvation potentials.** In this study, we use two models to investigate pressure-dependencies. The first model, termed "the CHOGG Model" (for Cheung-Hummer-Onuchic-Garcia-Gasic), is based off of work from Hummer and coworkers[17,18], which is derived from an information theory, and later expressed for off-lattice simulations[32,48] of protein folding under high pressure.

The information theory model accounts for the association, solvation, and conformational equilibria of hydrophobic solutes, such as residues in the core of a protein. These calculations are described by the probability of hydrophobic solutes forming a cavity in aqueous solvent. As a result of increasing pressure, the desolation barrier increases because of the increased free energy cost of forming a small cavity between two solutes. An information theoretic model will not encounter the usual systematic errors due to over-fitting of incorrect simulation data. Based on the Hummer *et al.*'s information theory calculation[17,18] and Hillson's simulation[49], Cheung and coworkers[32,48] created a desolvation barrier for structure-based models.

The second model for the investigation of pressure-dependencies, termed "the DC Model" (for Dias-Chan), is motivated by work from Dias and Chan[19] that calculate the potential of mean force between to methane molecules directly from simulation data.

The values of $\epsilon$, $\epsilon'$, and $\epsilon''$ are related to the magnitude of pressure by the following,

$$\epsilon = \epsilon_0 + \upsilon P + \xi P^2 \quad (5a)$$
$$\epsilon' = \epsilon_0' + \upsilon' P \quad (5b)$$
$$\epsilon'' = \epsilon_0'' + \upsilon'' P + \xi'' P^2 \quad (5c)$$

where $\epsilon_0'$ and $\epsilon_0''$ are the water-mediated well energy and the barrier height at ambient $P$, respectively. The constants for both models are given in Table 1. The constants for the CHOGG Model are taken from Ref.[49], and the constants for the DC model are from fitting to the PMF's (from Ref.[19]) minima and barrier height vs. $P$. Noting the values in Table 1, the CHOGG Model only uses the linear $P$-dependencies unlike the DC Model that has second order $P$ terms.

These *P*-dependent energies result in two very different *P*-dependent desolvation potentials, as shown in Figure 3. With the CHOGG Model, the desovlation barrier increases and the free energy gap between the two minima tilts to favor the water-mediated contact as pressure increases, leading to water penetrating the hydrophobic core and unfolding of a protein. Whereas with the DC Model, the free energy gap between the two minima tilts to favor the contact well as pressure increases, leading to a stabilization of the protein.

**Table 1.** Values of constants in pressure-dependent contact well, water-mediated well, and barrier height energies.

| constants | CHOGG Model | DC Model |
|---|---|---|
| $\epsilon_0'\ [\epsilon_0]$ | 0.33 | 0.19 |
| $\epsilon_0''\ [\epsilon_0]$ | 1.33 | 0.3 |
| $\upsilon\ [\sigma^3]$ | -0.127 | 0.039 |
| $\xi\ [\sigma^6/\epsilon_0]$ | 0 | 0.007 |
| $\upsilon'\ [\sigma^3]$ | 0 | 0.042 |
| $\upsilon''\ [\sigma^3]$ | 0.211 | 0.03 |
| $\xi''\ [\sigma^6/\epsilon_0]$ | 0 | 0.007 |

**System set up and simulation details.** We performed all simulations using GROMACS[55] to integrate Langevin equations of motion at a low friction limit. To prepare the computational model of yeast phosphoglycerate kinase (PDB ID: 1QPG) for the GROMACS simulations, we used SMOG: Structure-based Models for Biomolecules[56] software (http://smog-server.org/). A complete description of the simulation set up and protocol can be found in the supplemental material of Ref.[32].

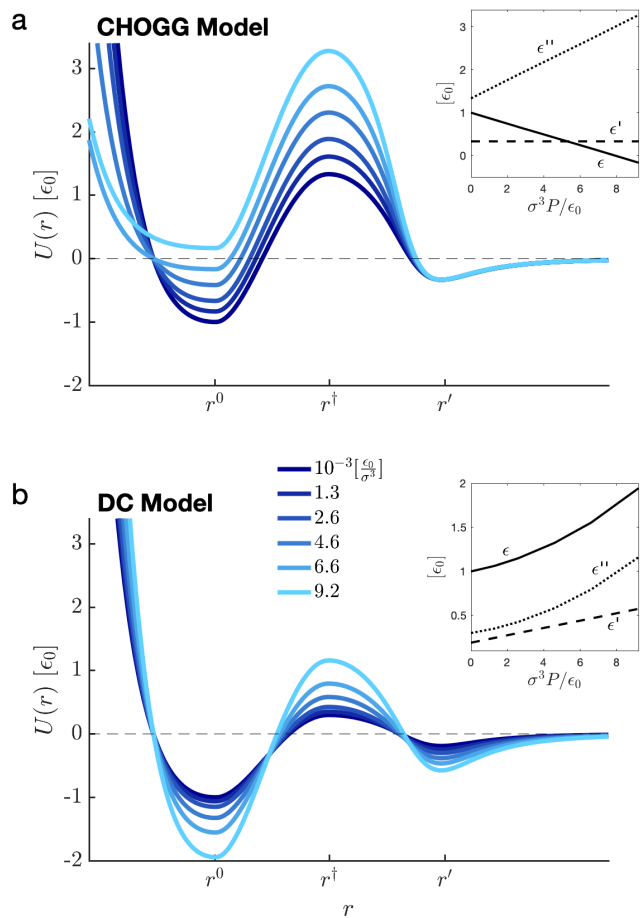

**Figure 3.** Pressure-dependent desolation PMF for between two residues at for varying $P$, including Lennard–Jones and solvent contributions. Results are shown for $\sigma^3 P/\epsilon_0 = $ 10-3 to 9.2 for (a) the CHOGG Model and (b) the DC Model. Insets show values of $\epsilon$, $\epsilon'$, and $\epsilon''$ as a function of $P$ (in reduced units, where $\epsilon_0/\sigma^3 \approx$ 76 MPa).

## Results

**The changes in the features of the desolvation potential will not impact the overall thermodynamic signatures at ambient $P$.** At ambient $P$ (= $10^{-3}$ $\epsilon_0/\sigma^3$), both models have the same contact well depth, resulting in similar unfolding curves when increasing $T$. Figure 4 shows that PGK unfolds at $T = T_f \approx 1.09$ for both models, which is indicated by the sudden increase in the average radius of gyration $R_g$ (normalized by the $R_g$ of the crystal structure, $R_g^0$). Here, $T_f$ is the temperature at which the protein folds or unfolds when cooling or heating, respectively. Note that the unfolded phase of the protein at high $T$ has a slightly lower $R_g$ for the DC Model than of the CHOGG Model. This minor difference in average $R_g$ maybe due to the DC Model having a lower desolvation barrier than that of the CHOGG Model, contributing to less water-mediated repulsion.

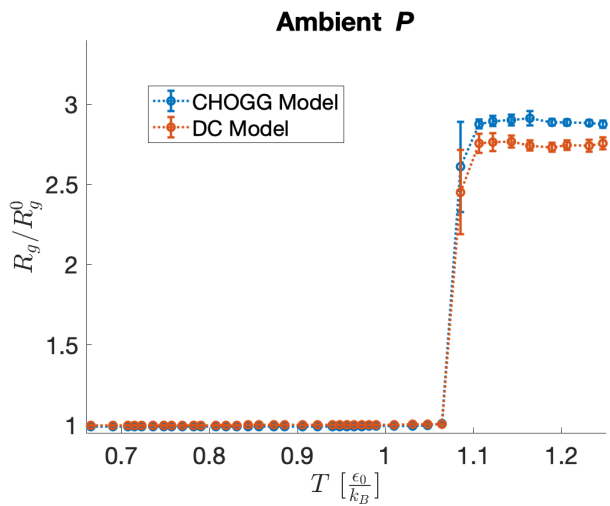

**Figure 4.** Normalized average $R_g$ as a function of $T$ for the CHOGG and DC Models at ambient $P$ (= $10^{-3}$ $\epsilon_0/\sigma^3$). Error bars are calculated using jackknife error estimation. $R_g^0$ is the radius of gyration of the crystal structure.

**Only the CHOGG Model captures protein denaturation at high pressure.** To understand the effects of high pressure on PGK, we examine two temperatures, one at $T = 0.87 < T_f$ (ambient $T$) and one at $T = 1.14 > T_f$ (high $T$). Before pressure is applied, both models result in an unfolded PGK above $T_f$ and folded PGK below $T_f$.

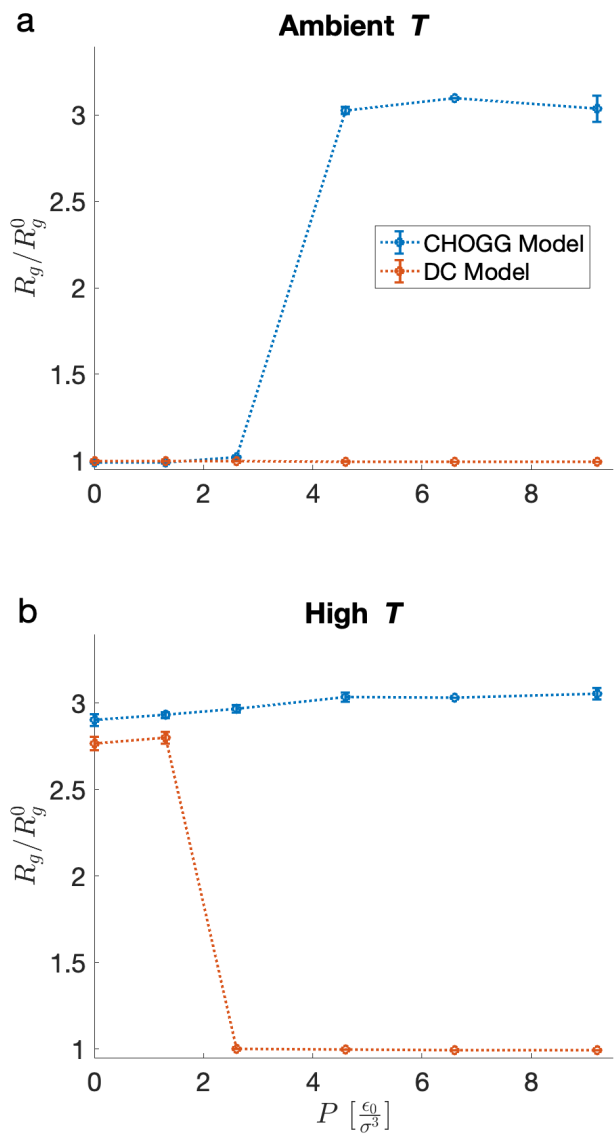

**Figure 5.** Normalized average $R_g$ as a function of $P$ for Models A and B at (a) ambient $T$ and (b) high $T$. Error bars are calculated using jackknife error estimation.

The average $R_g$ with respect to $P$ at ambient $T$, shown in Figure 5(a), demonstrates that the protein unfolds with the CHOGG Model and remains folded with the DC Model as pressure increases. Since the contact well strengthens with increased $P$ in the DC Model, the protein remains folded. Whereas, the CHOGG Model's contact well becomes weaker with increased $P$, causing the protein to unfold when $\epsilon < \epsilon'$ at $P = 4.6\ \epsilon_0/\sigma^3$. For reference, this pressure of $4.6\ \epsilon_0/\sigma^3$ is approximately equivalent to 350 MPa.

The effects of the contact wells strengthening (DC Model) or weakening (CHOGG Model) is also illustrated at high $T$. In Figure 5(b), as pressure increases, the protein folds with the DC Model, unlike the CHOGG Model where the protein remains unfolded. These two trends at different temperatures show that the CHOGG Model destabilizes the protein at high pressures, while the DC Model increases the proteins stability.

**The phase behavior of the two desolvation models.** Bringing the results together from Figures 4 & 5, we construct a $P$-$T$ phase diagram for both models, shown in Figure 6. The black line is the coexistence curve separating the folded and unfolded phases, and on the curve, both phases are equally populated due to having zero change in free energy ($\Delta G = 0$). The stability dependence on $P$ and $T$ is shown by the slope of the coexistence curve on the $P$-$T$ phase diagram. The negative slope, in Figure 6(a), shows that both increasing $P$ and $T$ destabilizes the protein with the CHOGG Model; whereas, the positive slope, in Figure 6(b), shows that shows that only increasing $T$ destabilizes the protein with the DC Model. From Eqn. (1) the slope of the coexistence curve is $\frac{\Delta S}{\Delta V}$. Because both models have a positive $\Delta S$ (entropy increases upon unfolding), therefore $\Delta V$ must be negative for the CHOGG Model and positive for the DC Model. Thus, CHOGG model captures the protein denaturation under pressure.

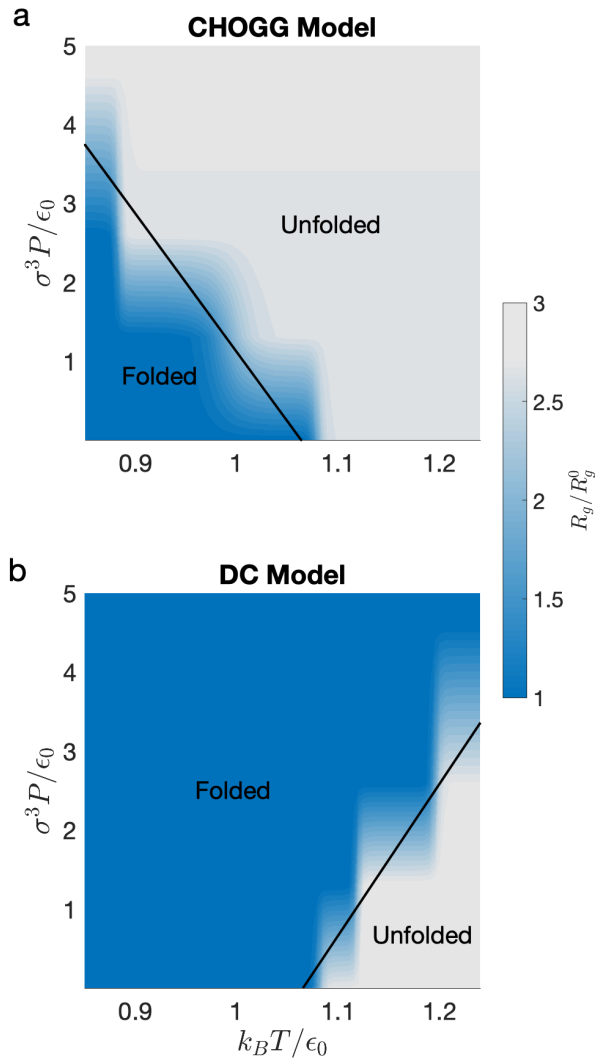

**Figure 6.** *P-T* phase diagram for (a) the CHOGG Model and (b) the DC Model. The blue region corresponds to compact values of $R_s$ and the gray region corresponds to extended values of $R_s$, which signify folded and unfolded phases, respectively. The black line is a linear approximation of the phase boundary between folded and unfolded protein phases.

From Figure 6, the slopes of the black line indicate that $\Delta V < 0$ for the CHOGG Model (Figure 6(a)) and $\Delta V > 0$ for the DC Model (Figure 6(b)). The behavior on the phase diagram in Figure 6(a) for the CHOGG Model may occur in real proteins if the $\Delta V = 0$ line is shifted to a low or negative pressure and the slope is flattened similar to the orange curve in Figure 1 (to the right of the $\Delta S = 0$ line). Whereas, the phase diagram of the DC Model in Figure 6(b) is similar to the blue curve in Figure 1 at high temperature below the $\Delta V = 0$ line (i.e. $\Delta V > 0$ ).

## Discussion

**The CHOGG Model captures protein denaturation under pressure unlike the DC model.** Both the CHOGG and the DC models do not capture the full range of *P-T* plane from Hawley's theory; however, our simulation results conclusively show that the CHOGG model is best suited for studying hydrostatic pressure unfolding when in the $\Delta V < 0$ regime, which the DC Model does not acutely reflect the phenomenological behavior.[32] Indeed, the CHOGG model accurately describes the unfolding behavior of PGK under high pressure, including the nontrivial existence of a stable intermediate at low *T* and high *P*. The phase diagrams created by the CHOGG model was validated by experiments and by an analytical theory.[32] The DC Model is by no means incorrect. The DC Model may be more suited for studying the effects of pressure in the $\Delta V > 0$ regime such as an $\alpha$-helix peptide, which has a smaller volume folded than unfold. By combining aspects of both models, an elliptical coexistence curve can be achieved with the desired center and $\Delta V$, $\Delta S$ axes rotation or shearing as seen in Figure 1.

**The PMFs of methane molecules may not be fully representative of the PMFs for proteins.** These two PMFs (Figure 3(a) and (b)) may not be in conflict with each other though, because temperature will affect the pressure-dependence as well. The change in the contact well depth, whether it increases or decreases, depends on the temperature, which is shown by Ashbaugh *et al.*[20] through calculating the second virial coeficient. They show that at high *T*, the second virial coefficient increases (less attraction) as pressure increases, while at low *T*, second virial coefficient decreases (more attraction) as pressure increases. The pressure-dependence of the contact well of the PMF between the methanes can be interpreted as becoming shallower or deeper as pressure increases at high *T* or low *T*, respectivly, which is individually captured by the CHOGG or DC models.

Furthermore, using perfectly detailed PMFs from two methane molecules as the pairwise hydrophobic interaction of a protein will not be the "true" PMF of the protein for two reasons: i.) simply, protein residues are not methane molecules, and ii.) many-body and emergent effects of having many hydrophobic molecules together will change the PMF. For example, the PMF between two hydrophobic plates has a different pressure-dependence than that of two methane molecules.[57] For the PMF between the graphene plates immersed in TIP4P/2005 water at $T = 300$ K, as pressure increases, the contact well initially deepens until 800 MPa and becomes shallower at 1200 MPa.

Another example showing where the PMF of methane molecules contradicts the PMF of proteins: Dias also shows a different pressure-dependence in the PMF of a simple protein-water model[27] than that of the two methane molecules that he and Chan propose in Ref.[19] (the DC Model in this study). The PMF of a simple protein-water model from Dias's work[27] is comparable to Figure 3(a) (the CHOGG Model), which opposes the trend of Figure 3(b) (the DC Model). The protein-water model from Dias[27] and the CHOGG both have a contact well the weakens as pressure increases, whereas in the DC model, the contact well strengthens with increased pressure. Dias and Chan even state that proteins will have a different pressure-dependent hydrophobic interaction than that of methane's:

Conceptually, however, it is important to recognize that two- and three-methane PMFs do not, by themselves, necessarily provide an adequate physical picture of pressure denaturation because the two- and three-body contact minima retain significant water exposure. Hence, the adequacy of these configurations as models for the sequestered folded protein core can be limited.[19]

In their conclusion of their paper,[19] they also describe the difficulty of correlating the combined pressure and temperature dependencies of the methane PMF to that of real proteins.

**The merit of coarse-grained modelling over all-atom simulations.** Structure-based coarse-grained models have a funneled energy landscape with minimal frustration.[12,13] As we learned from the energy landscape theory of protein folding,[58] structure-based models provide fruitful insights. Why is that?—because, the funneled energy landscape itself is emergent and microscopic details cease to matter.[59] This is evident with mutation experiments; few perturbations of the residue sequence retain the same topological structure. Thus, the exact molecular-scale details and chemistry are not as important as the essential physics provided by structure-based models.

Coming back to the problem at hand, regardless of the exact details, the essential physics of the CHOGG Model is that the desolvation barrier increases and the free energy gap between the two minima tilts to favor the water-mediated contact as pressure increases, leading to water penetrating the hydrophobic core and unfolding of a protein. This is captured by the PMF calculated by Hummer and coworkers,[17,18,60] and has been used by others to understand pressure-denaturation.[40,49,61–64] In our recent study in collaboration with experimentalists (which inspired this work),[32] we capture important thermodynamic trends in the pressure-denaturation of PGK. These trends include the evidence for critical behavior of a protein (note that criticality is another well-known emergent phenomenon[65]; not discussed here), which is not explicit in our model.

## Conclusion

Through our investigation, we have compared two different pressure-dependent desolvation potentials and how the two models may be used in structure-based minimalist models to further understand the full *P-T* phase diagrams of real proteins, as intended by Hawley's original work.

AUTHOR INFORMATION

**Corresponding Author**

*Email: mscheung@uh.edu

**Author Contributions**

A.G.G. ran the simulations, and both authors designed the project and wrote the paper.

**Notes**

The authors declare no competing financial interest.

ACKNOWLEDGMENT

We thank the computing resources from the Hewlett Packard Enterprise Data Science Institute at UH. A.G.G. was supported by a training fellowship on the Houston Area Molecular Biophysics Program (T32 GM008280). We also thank funding from the National Science Foundation (MCB-1412532, OAC-1531814). Lastly, the inspiration for writing this paper stems from an anonymous reviewer's critique of our previous work—for that, we are grateful.

ABBREVIATIONS

PMF, potential of mean force;

Table of Contents graphic

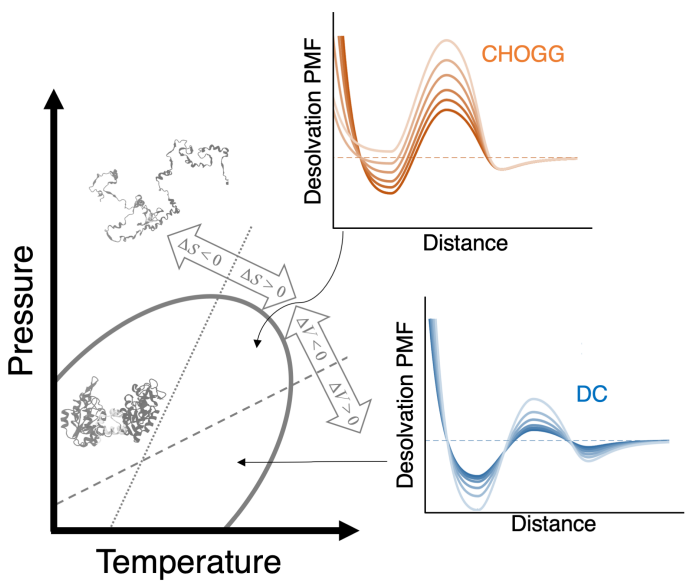